# Frenet-Serret analysis of helical Bloch modes in N-fold rotationally symmetric rings of coupled spiralling optical waveguides

Y. CHEN[1,*] AND P. ST.J. RUSSELL[1,2]

[1]*Max Planck Institute for the Science of Light and* [2]*Department of Physics, Friedrich-Alexander-Universität, Staudtstr. 2, 91058 Erlangen, Germany*
*\*yang.chen@mpl.mpg.de*

**Abstract:** The behavior of electromagnetic waves in chirally twisted structures is a topic of enduring interest, dating back at least to the invention in the 1940s of the microwave travelling-wave-tube amplifier and culminating in contemporary studies of chiral metamaterials, metasurfaces, and photonic crystal fibers (PCFs). Optical fibers with chiral microstructures, drawn from a spinning preform, have many useful properties, exhibiting for example circular birefringence and circular dichroism. It has recently been shown that chiral fibers with *N*-fold rotationally symmetric (symmetry group $C_N$) transverse microstructures support families of helical Bloch modes (HBMs), each of which consists of a superposition of azimuthal Bloch harmonics (or optical vortices). An example is a fiber with *N* coupled cores arranged in a ring around its central axis (*N*-core single-ring fiber). Although this type of fiber can be readily modelled using scalar coupled mode theory, a full description of its optical properties requires a vectorial analysis that takes account of the polarization state of the light—particularly important in studies of circular and vortical birefringence. In this paper we develop, using an orthogonal two-dimensional helicoidal coordinate system embedded in a cylindrical surface at constant radius, a rigorous vector coupled mode description of the fields using local Frenet-Serret frames that rotate and twist with each of the *N* cores. The analysis places on a firm theoretical footing a previous HBM theory in which a heuristic approach was taken, based on physical intuition of the properties of Bloch waves. After a detailed review of the polarization evolution in a single spiralling core, the analysis of the *N*-core single-ring system is carefully developed step by step. The accuracy limits of the analysis are assessed by comparison with the results of finite element modelling, focusing in particular on the dispersion, polarization states and transverse field profiles of the HBMs. We believe this study provides clarity in what can sometimes be a rather difficult field, and will facilitate further exploration of real-world applications of these fascinating waveguiding systems.

## 1. Introduction

Studies of the propagation of electromagnetic waves in helically twisting structures can be traced back to the 1940s, when there was considerable interest in helical antennas and twisted waveguides, and when Kompfner proposed the travelling wave tube—a microwave amplifier in which the electromagnetic wave is guided along a spiralling wire, allowing its axial velocity to be matched to that of a sub-luminal axial electron beam [1] [2]. Lewin analyzed the modes guided in twisted waveguides of rectangular cross-section [3] and Waldron proposed a helical coordinate system and applied it to a variety of different helical structures [4]. Interest in helical waveguides at optical frequencies emerged in the 1980s with the advent of single-mode optical fibers, which could be readily twisted and coiled. This was of particular relevance to magnetic field and current sensing by the Faraday effect, since mechanical twisting of a single mode fiber (SMF) suppressed unwanted residual linear birefringence, permitting undistorted transmission of circularly polarized light [5]. The first fibers drawn while spinning the preform ("spun fibers") were reported in 1981, the aim being to reduce residual linear birefringence [6]. The properties of helically coiled SMF were subsequently placed on a firm theoretical footing by Ross, who made use of a local Frenet-Serret frame to follow the field evolution along the spiral path [7] [8]. It was subsequently realized that linearly birefringent (HiBi) fibers drawn from a spun preform would support elliptically polarized eigenmodes, also of use in current sensing [9]. Experimental confirmation followed of strong circular birefringence in spun fiber with an off-centre core [10]. Menyuk predicted in 1994 that an elliptically birefringent fiber, when helically twisted by a precise amount, would allow elimination of Kerr-related nonlinear ellipse rotation [11].

The realization that photonic crystal fiber [12] could be drawn in chiral form has resulted in a resurgence of interest in chiral fibers [13] [14] [15] [16] [17] [18] [19], in particular those with multiple cores or complex transverse microstructures. Longhi provided one of the first theoretical studies of Bloch dynamics of light in helical waveguide arrays, using a scalar analysis to explore analogies with electrons in two-dimensional periodic potentials [20]. The group of Alexeyev has made many valuable theoretical contributions, including using a Frenet-Serret frame to analyze a helical fiber with a single off-axis core [21] and studies of optical activity in fibers with central helicoidal cores [22]. The structural complexity of helically twisted photonic crystal fibers made it necessary to develop finite element techniques in helicoidal coordinates, now in standard use for numerical experiments [23]. There has been much experimental work also, including observation of spectral dips in twisted PCF, caused by coupling to azimuthal resonances in the cladding [18] [24], use of off-axis spiralling cores to strip off higher order modes in both solid [25] and hollow core [26] PCFs, creation of strong circular dichroism in helically twisted hollow core PCF [27] and twist-induced guidance in a coreless PCF [28].

In this paper we develop a rigorous analytical model, based in the local Frenet-Serret frame, of helical Bloch modes guided in rotationally symmetric fibers with $N$ linearly birefringent cores arranged in a single ring around the axis, which we denote by the acronym N-SRF. The analysis, within the paraxial approximation, is fully vectorial and valid for the case when the birefringent axes of the cores are tilted relative to the radial and azimuthal directions. It remedies incomplete results from previous papers, especially in regard to field polarization, in the process clarifying the roles of azimuthal order, orbital angular momentum and spin-orbit coupling as well as describing the procedures needed to convert between the Frenet-Serret and the laboratory frames. We believe the resulting—rather simple—analytical results will make this difficult field more easily accessible to researchers working on future applications of helical fibers.

The paper is organized as follows. First we define an orthogonal system of helicoidal coordinates, and explain how it is related to Cartesian coordinates in the laboratory frame. Then we use this coordinate system to treat the case of a single birefringent spiralling core, using a paraxial model based on circularly polarized field components, which turns out to produce appealingly elegant analytical solutions. These single-core solutions are then used to construct, using vector coupled mode theory, the helical Bloch modes (HBMs) of a fiber with $N$ identical cores arranged in a ring about the axis. The analysis is then used to model a realistic 6-SRF structure and the results compared at each stage with those obtained by finite element modelling (FEM), which also allows us to estimate the twist-induced loss.

## 2. Parametric model of helical curve

We adapt and extend the physical picture developed in the 1980s for propagation in helically coiled waveguides [7]. Essentially, the modal fields are assumed to be locally paraxial and transverse to the axis of the spiralling core, within the Frenet-Serret frame, which rotates and twists with the spiral. To treat N-SRF, we introduce a coordinate system consisting of a mesh of orthogonal helical curves on the surface of a cylinder of constant radius $\rho$ (Fig. 1). As we shall later see, this system is especially convenient when modelling helical Bloch waves, which progress along the spiralling waveguides while the Bloch phase advances perpendicular to them.

We choose a parametrization that makes the derivatives of the position vector automatically unit vectors (see below):

$$x(\sigma_1,\sigma_2) = \rho\cos\phi, \quad y(\sigma_1,\sigma_2) = \rho\sin\phi, \quad z(\sigma_1,\sigma_2) = \frac{\sigma_1 - \sigma_2\alpha\rho}{\sqrt{1+\rho^2\alpha^2}} \tag{1}$$

$$\phi = (\alpha\sigma_1 + \sigma_2/\rho)/\sqrt{1+\rho^2\alpha^2}$$

where $\sigma_1$ and $\sigma_2$ are the orthogonal helicoidal coordinates and $\alpha$ is the twist rate of the $\sigma_2 =$ constant curves. The $(\sigma_1,\sigma_2)$ coordinates can be readily converted to Cartesian cylindrical coordinates $(\phi, z)$:

$$\begin{pmatrix} \rho\phi \\ z \end{pmatrix} = \begin{pmatrix} 1/\sqrt{1+\rho^2\alpha^2} & \alpha\rho/\sqrt{1+\rho^2\alpha^2} \\ -\alpha\rho/\sqrt{1+\rho^2\alpha^2} & 1/\sqrt{1+\rho^2\alpha^2} \end{pmatrix} \begin{pmatrix} \sigma_2 \\ \sigma_1 \end{pmatrix} = [\mathbf{M}_C]\begin{pmatrix} \sigma_2 \\ \sigma_1 \end{pmatrix} \tag{2}$$

which shows as expected that the two coordinate systems are linked by rotation through the angle $\sin^{-1}(\alpha\rho/\sqrt{1+\alpha^2\rho^2})$.

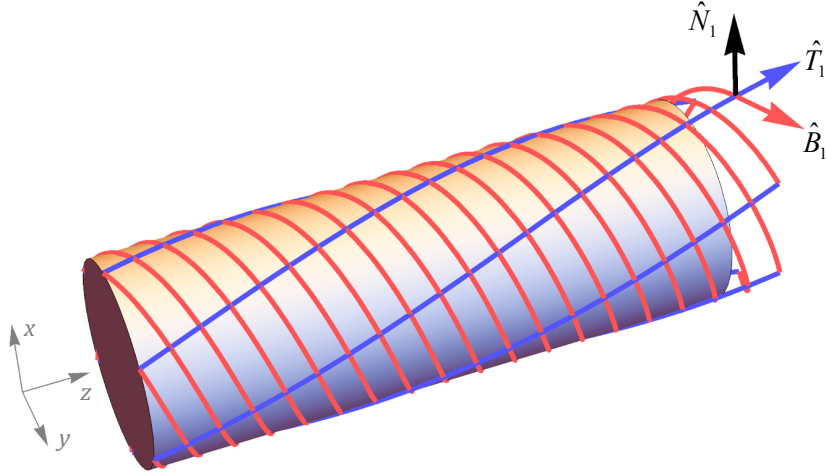

Fig. 1: Sketch of the orthogonal helicoidal coordinate system used in the paper, for twist rate $\alpha < 0$, on a surface of constant radius. In a twisted 6-SRF the cores would be aligned along the blue curves, while the Bloch phase evolves along the orthogonal red curves. Note the unit vectors for the blue curves, in the normal ($\hat{N}_1$, black), tangential ($\hat{T}_1$, blue), and binormal ($\hat{B}_1$, red) directions (see text).

Equation (1) yields tangential, normal and binormal unit vectors $(\hat{T}_1, \hat{N}_1, \hat{B}_1)$ (for the $\sigma_2 =$ constant curves), which are simply related to $(\hat{T}_2, \hat{N}_2, \hat{B}_2)$ (for the $\sigma_1 =$ constant curves):

$$\hat{T}_1 = -\hat{B}_2 = \frac{\partial \mathbf{r}}{\partial \sigma_1} = (-\alpha\rho\sin\phi, \alpha\rho\cos\phi, 1)/\sqrt{1+\rho^2\alpha^2}$$

$$\hat{N}_1 = \hat{N}_2 = \hat{B}_1 \times \hat{T}_1 = \hat{B}_2 \times \hat{T}_2 = (\cos\phi, \sin\phi, 0) \qquad (3)$$

$$\hat{B}_1 = \hat{T}_2 = \frac{\partial \mathbf{r}}{\partial \sigma_2} = (-\sin\phi, \cos\phi, -\alpha\rho)/\sqrt{1+\rho^2\alpha^2}.$$

If the twist rate $\alpha$ is positive, the $\sigma_2 = $ constant curves rotate clockwise, viewed in the $+\sigma_1$ direction, with curvature:

$$\kappa_1 = \frac{1}{R_1} = \left\| \frac{d\hat{T}_1}{d\sigma_1} \right\| = \frac{\alpha^2 \rho}{1+\alpha^2\rho^2} \qquad (4)$$

and torsion $\tau_1$:

$$\frac{\partial \hat{B}_1}{\partial \sigma_1} = \frac{\partial \hat{B}_1}{\partial \phi}\frac{\partial \phi}{\partial \sigma_1} = \frac{\alpha}{1+\rho^2\alpha^2}(-\cos\phi, -\sin\phi, 0) = -\tau_1 \hat{N}_1. \qquad (5)$$

$\hat{T}_1$, $\hat{N}_1$ and $\hat{B}_1$ in the above equations satisfy the Frenet-Serret equations, as required:

$$\frac{d\hat{T}_1}{d\sigma_1} = -\kappa_1 \hat{N}_1, \quad \frac{d\hat{N}_1}{d\sigma_1} = \kappa_1 \hat{T}_1 + \tau_1 \hat{B}_1, \quad \frac{d\hat{B}_1}{d\sigma_1} = -\tau_1 \hat{N}_1. \qquad (6)$$

Other relations that we will use are:

$$\frac{d^2\hat{N}_1}{d\sigma_1^2} = -(\kappa_1^2 + \tau_1^2)\hat{N}_1, \quad \frac{d^2\hat{B}_1}{d\sigma_1^2} = -\kappa_1\tau_1\hat{T}_1 - \tau_1^2\hat{B}_1, \quad \frac{d^2\hat{T}_1}{d\sigma_1^2} = -\kappa_1^2\hat{T}_1 - \kappa_1\tau_1\hat{B}_1. \qquad (7)$$

Note that the orthogonal set of curves ($\sigma_1 = $ constant) has parameters:

$$\alpha_2 = -\alpha^{-1}\rho^{-2}, \quad \kappa_2 = \frac{1/\rho}{1+\alpha^2\rho^2} = \frac{\kappa_1}{\alpha^2\rho^2}, \quad \tau_2 = \frac{\alpha}{1+\alpha^2\rho^2} = \tau_1. \qquad (8)$$

### 3. Single off-axis core

Before treating the system of coupled cores, it is important first to analyze the properties of a single spiralling core. We treat the structure illustrated in Fig. 2(a), where a single linearly birefringent core, tilted at an angle $\phi_T$ to the radial direction, is placed at distance $\rho$ from the axis. The core is assumed to be single-mode, i.e., it supports only fundamental fast and slow modes, with $E$-fields polarized along the $F$ and $S$ axes, respectively.

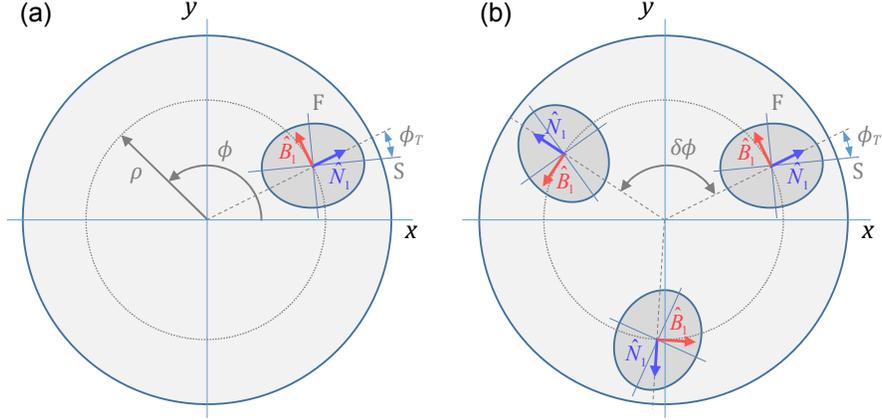

Fig. 2: Example cross-sections of the structures considered in the paper, pictured at an arbitrary position along the fiber (note that $+z$ points out of the paper). The angle $\phi = \alpha z$ increases in the direction of the spiral. $\hat{N}_1$ and $\hat{B}_1$ are the normal and binormal unit vectors in the Frenet-Serret frame, within which the fast (F) and slow (S) birefringent axes of the cores are tilted at an angle $\phi_T$. For $\alpha > 0$ $\hat{B}_1$ tilts slightly into the $-z$ direction (see Fig. 1), and the tangent vector $\hat{T}_1 = \hat{N}_1 \times \hat{B}_1$ points along the spiralling cores in the direction of propagation. (a) A single core structure and (b) a multi-core rotationally symmetric structure with $N = 3$.

## 3.1 Eigenmodes of spiraling core

The trajectory of a single spiraling core is given by one of the curves for $\sigma_2 =$ constant and can be described by:

$$x(\sigma_1) = \rho \cos(\alpha' \sigma_1 + \phi_0),\ y(\sigma_1) = \rho \sin(\alpha' \sigma_1 + \phi_0),\ z(\sigma_1) = \frac{\sigma_1}{\sqrt{1+\alpha^2 \rho^2}} - \phi_0 \alpha \rho^2 \quad (9)$$

where $\alpha' = \alpha / \sqrt{1+\alpha^2 \rho^2}$ and $\phi_0$ is the angular position of the core at $\sigma_1 = 0$. The analysis turns out to be much more elegant if the field in the core is expanded as a sum of LCP and RCP components, i.e.:

$$\begin{aligned}\mathbf{E}(\sigma_1) &= e_L(\sigma_1) \frac{\hat{N}_1 + i\hat{B}_1}{\sqrt{2}} e^{-i\phi_T} e^{i\beta_0 \sigma_1} + e_R(\sigma_1) \frac{\hat{N}_1 - i\hat{B}_1}{\sqrt{2}} e^{i\phi_T} e^{i\beta_0 \sigma_1} \\ &= e_L(\sigma_1) \hat{u}(\sigma_1) e^{i\beta_0 \sigma_1} + e_R(\sigma_1) \hat{u}^*(\sigma_1) e^{i\beta_0 \sigma_1}\end{aligned} \quad (10)$$

where $\hat{u}(\sigma_1) = [(\hat{N}_1 + i\hat{B}_1)/\sqrt{2}] \exp(-i\phi_T)$ is a complex-valued unit vector. The core tilt angle $\phi_T$ within the Frenet-Serret frame adds to the phase of the RCP field and subtracts from the phase of the LCP field, causing the axes of the polarization ellipse to line up with the birefringent axes of the core. The slow and fast modes in the core have electric fields oscillating along the unit vectors $\hat{S} = \hat{N}_1 \cos\phi_T - \hat{B}_1 \sin\phi_T$ and $\hat{F} = \hat{N}_1 \sin\phi_T + \hat{B}_1 \cos\phi_T$, with amplitudes related to $e_L$ and $e_R$ by:

$$\begin{pmatrix} e_S \\ e_F \end{pmatrix} = \frac{1}{\sqrt{2}} \begin{pmatrix} 1 & 1 \\ i & -i \end{pmatrix} \begin{pmatrix} e_L \\ e_R \end{pmatrix}. \quad (11)$$

The total field satisfies the wave equation in the Frenet-Serret frame:

$$\frac{d^2}{d\sigma_1^2}\left((\hat{S}e_S + \hat{F}e_F)e^{i\beta_0\sigma_1}\right) + (\beta_S^2\hat{S}e_S + \beta_F^2\hat{F}e_F)e^{i\beta_0\sigma_1} = 0, \tag{12}$$

where $\beta_S$ and $\beta_F$ are the propagation constants in an isolated, untwisted, linearly birefringent core, with:

$$\beta_S = \beta_0 + \vartheta/2, \quad \beta_F = \beta_0 - \vartheta/2 \tag{13}$$

where $\beta_0 = kn_0$ is the mean propagation constant, $k$ is the vacuum wavevector and $\vartheta = \beta_S - \beta_F$. Substituting Eqs. (10) and (11) into Eq. (12) and rearranging leads to:

$$\frac{d^2}{d\sigma_1^2}\left([e_L\hat{u} + e_R\hat{u}^*]e^{i\beta_0\sigma_1}\right) + \frac{1}{\sqrt{2}}\left((\beta_S^2\hat{S} - i\beta_F^2\hat{F})e_R + (\beta_S^2\hat{S} + i\beta_F^2\hat{F})e_L\right)e^{i\beta_0\sigma_1} = 0. \tag{14}$$

Upon taking the scalar product with $\hat{u}$ and $\hat{u}^*$, we obtain two coupled equations for the amplitudes $e_R$ and $e_L$. The differential operator in Eq. (14) produces a series of derivatives of the unit vectors, each of which has a different scalar product with $\hat{u}$:

$$\begin{aligned}
\hat{u}\cdot\hat{u}^* &= 1, & \hat{u}\cdot\hat{u} &= 0, & \hat{u}\cdot\hat{u}'^* &= i\tau, \\
\hat{u}\cdot\hat{u}'''^* &= -\kappa^2/2 - \tau^2, & \hat{u}\cdot\hat{S} &= 1/\sqrt{2}, & \hat{u}\cdot\hat{F} &= i/\sqrt{2}.
\end{aligned} \tag{15}$$

The scalar products with $\hat{u}^*$ are simply the complex conjugates of these relations. Evaluating the derivatives in Eq. (14) and taking its dot product with $\hat{u}$ and $\hat{u}^*$, neglecting second order derivatives of $e_R$ and $e_L$ and noting that for our parameters $|\tau/\beta_0| \ll 1$, $|\vartheta/\beta_0| \ll 1$, and that $\kappa^2$ is orders of magnitude smaller than the other terms, we obtain the eigensystem:

$$\begin{pmatrix} \gamma - \tau & -(\vartheta/2)e^{i2\phi_T} \\ -(\vartheta/2)e^{-i2\phi_T} & \gamma + \tau \end{pmatrix}\begin{pmatrix} e_L \\ e_R \end{pmatrix} = 0 \tag{16}$$

where $i\gamma = \partial/\partial\sigma_1$. Analysis of Eq. (16) yields the eigen-solutions:

$$\gamma_\pm = \pm\sqrt{\tau^2 + \vartheta^2/4}, \quad (e_{L\pm}, e_{R\pm}) = \left(e^{i2\phi_T}(\tau \pm \sqrt{\tau^2 + \vartheta^2/4}), \vartheta/2\right)$$

where in the limit as $\vartheta \to 0$ the fields become uncoupled (Eq. (16) becomes diagonal) and the plus sign in the eigenvalue yields an LCP mode, and vice-versa. We can write the total eigen-field as:

$$\mathbf{E}_\pm = (e_{L\pm}\hat{u} + e_{R\pm}\hat{u}^*)e^{i(\beta_0+\gamma_\pm)\sigma_1} = \left(\frac{e_{L\pm}e^{-i\phi_T} + e_{R\pm}e^{i\phi_T}}{\sqrt{2}}\hat{N}_1 + i\frac{e_{L\pm}e^{-i\phi_T} - e_{R\pm}e^{i\phi_T}}{\sqrt{2}}\hat{B}_1\right)e^{i(\beta_0+\gamma_\pm)\sigma_1} \tag{17}$$

which shows that the polarization state (in general elliptical) is constant within the spiralling Frenet-Serret frame, while in the laboratory frame the polarization ellipse rotates with the fiber twist. For circularly polarized eigenmodes, i.e., $\vartheta = 0$, Eq. (17) yields a particularly simple result:

$$\mathbf{E}_\pm = e_0(\hat{N}_1 \pm i\hat{B}_1)e^{i(\beta_0\pm\tau)\sigma_1} = e_0 e^{i\phi_0}(\hat{x} \pm i\hat{y})e^{i(\beta_0\pm(\tau-\alpha'))z\sqrt{1+\alpha^2\rho^2}} \tag{18}$$

where $e_0$ is an arbitrary field amplitude, the approximation $\sqrt{1+\alpha^2} \simeq 1$ has been used, and the very small $z$-component has been neglected. This gives a residual circular birefringence in the laboratory frame of:

$$B_C = \left| \frac{\lambda}{\pi}(\tau_1 \sqrt{1+\alpha^2\rho^2} - \alpha) \right| \simeq \frac{\rho^2 \alpha^3 \lambda}{2\pi} \tag{19}$$

as has been previously reported [7] [8].

In the special case of a ring of non-birefringent cores in a twisted photonic crystal fiber, the modes in each core are $N$-fold rotationally symmetric, with the result that they are forced to rotate with the twist, resulting in additional circular birefringence [19]. Depending on the twist rate, this can act to either enhance or reduce the circular birefringence in Eq. (19); indeed, at a special value of twist rate the overall circular birefringence goes to zero. We do not take account of this effect in this paper.

For $\vartheta \neq 0$ the eigen-fields are in general elliptically polarized in the helicoidal frame, while in the laboratory frame the plane of polarization rotates through α radians per meter along the fiber. The behavior is governed by the relative strengths of the torsion and the linear birefringence, quantified by $2\tau_1/\vartheta$. For $2\tau_1/\vartheta \gg 1$, a linearly polarized incident field will almost be able to "ignore" the twist, being only slightly rotated as it travels along the fiber. For $2\tau_1/\vartheta \ll 1$ on the other hand, a field that is linearly polarized and aligned with unit vectors $\hat{S}_1$ or $\hat{F}_1$ will follow the twist, i.e., the fiber will preserve linear polarization state.

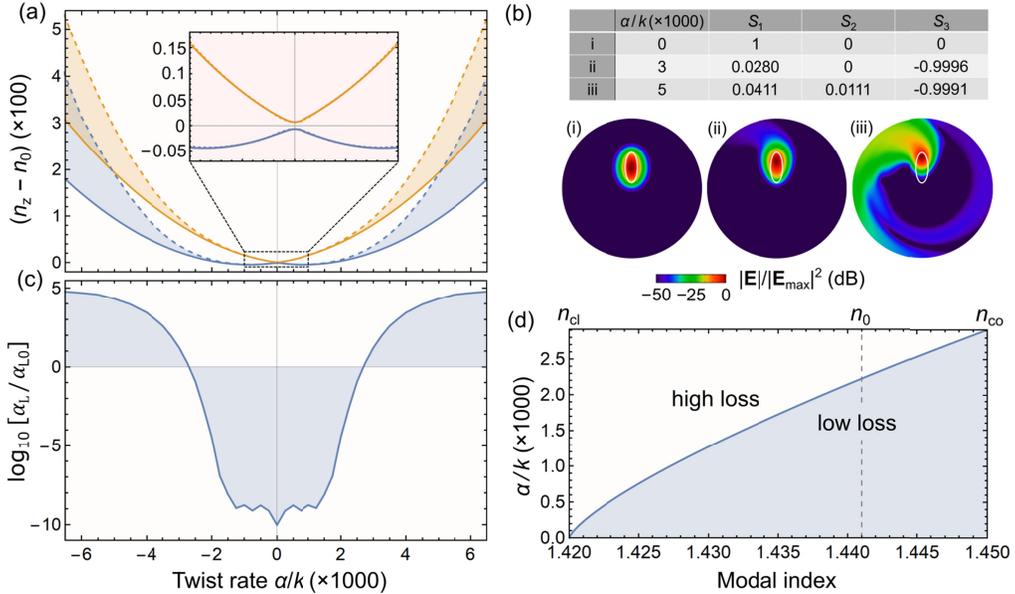

Fig. 3: Finite element modelling (FEM) of the twist-dependence of the modal parameters in a single-core fiber with $k\rho_{ca} = 6.5\pi$, $k\rho_{cb} = 3.2\pi$ and $k\rho = 9\pi$, where $k$ is the vacuum wavevector. The core and cladding indices are 1.45 and 1.42. See text for other parameters. (a) Comparison of the axial refractive indices of the two modes in the laboratory frame calculated by FEM (dashed curves) and using the analytical model (full curves). The inset is a zoom into behavior at low twist rates, highlighting the small anti-crossing at zero twist rate caused by the linear birefringence of the core. The analytical Stokes parameters are in almost perfect agreement with the values calculated by FEM (less than 1.5% error at all twist rates) so are not shown. (b) Distributions of the electric field intensity of the modes at twist rates $\alpha/k = 0$, 0.003 and 0.005, calculated by FEM. As the twist rate rises, the mode peak moves outwards and the polarization state becomes almost perfectly circular. At the highest twist rate the light leak out strongly from the core. (c) Modal loss $\alpha_L$ as a function of twist rate, calculated by FEM. Zero on this plot corresponds to $\alpha_L = \alpha_{L0} = 1$ dB/m. The loss is almost identical for both modes (only one is shown). (d) Plot of $0.2|\alpha/k|_{crit}$ versus modal index. For $|\alpha/k|$ values greater than $\sim 0.2|\alpha/k|_{crit}$ the leakage loss due to the waveguide curvature grows strongly (see text).

*3.2 Comparison with numerical modelling*

To explore the accuracy of this single-core analysis, we now compare its results with those from full vectorial finite element modelling (FEM) of a realistic fiber, based on a helicoidal coordinate system [23]. The structure analyzed has core and cladding indices 1.45 and 1.42, a single elliptical core (halfwidth $\rho_{ca}$ along the major axis and $\rho_{cb}$ along the minor axis), and normalized parameters $k\rho_{ca} = 6.5\pi$, $k\rho_{cb} = 3.2\pi$ and $k\rho = 9\pi$. For these parameters, FEM yields, for an isolated untwisted core, $n_0 = 1.441$ and $\vartheta/k = 0.00014$, which are the values used in the analytical model. A further advantage of FEM is that it allows us to estimate the bend loss induced by the spiral path [29].

The axial index of the two eigenmodes in the laboratory frame, given analytically by:

$$n_z = \left((\beta_0 \pm \sqrt{\tau^2 + \vartheta^2/4})\sqrt{1+\alpha^2\rho^2}\right)/k \tag{20}$$

is compared with the results of FEM in Fig. 3(a). It is noticeable that the FEM results increasingly deviate from the analytical values for twist rates $|\alpha/k| > \sim 0.002$. The reason for this is clear from the field profiles (Fig. 3(b)), which show that the mode is centrifugally thrown outwards in the twisted fiber, increasing the effective value of $\rho$ and thus the axial index, so that it no longer matches the analytical result. The modal polarization states in the local Frenet-Seret frame calculated by FEM and the analytical model are in almost perfect agreement, and are very close to circular for $|\alpha/\vartheta| \gg 1$ (Fig. 3(b)).

The leakage loss calculated by FEM (Fig. 3(c)) shows a marked dependence on twist, reaching as high as $10^5$ dB/m at $|\alpha/k| = 0.006$, but remaining below 1 dB/m for twist rates $|\alpha/k| < \sim 0.002$. At the highest twist rate light is clearly ejected in a curved stream from the twisting core, reminiscent of the trail of a comet following a curved path (Fig. 3(b)). An analytic estimate of the twist rate at which bend loss becomes significant can be obtained by substituting the radius of curvature (from Eq. (4)) into the well-known expression for the critical bend radius $R_{crit}$ [30]:

$$\frac{1+\alpha^2\rho^2}{\alpha^2\rho} \gg R_{crit} = \frac{2n_{co}^2}{k(n_0^2 - n_{cl}^2)^{3/2}} \tag{21}$$

where $n_{co}$ and $n_{cl}$ are the core and cladding indices, and $n_0$ is the modal index. Making the approximation that $\alpha^2\rho^2 \ll 1$, we obtain the condition:

$$|\alpha/k| \ll |\alpha/k|_{crit} = \sqrt{\frac{(n_0^2 - n_{cl}^2)^{3/2}}{2n_{co}^2 k\rho}} \tag{22}$$

for low loss. A plot of $\alpha/k = 0.2 \times |\alpha/k|_{crit}$ versus $n_0$ for $k\rho = 9\pi$, $n_{co} = 1.45$ and $n_{cl} = 1.42$ is shown in Fig. 3(d). For $n_0 = 1.441$ (the value used in the FEM), this predicts that the loss will grow strongly at twist rates $\alpha/k > \sim 0.002$, which is in reasonable agreement with the FEM results in Fig. 3(a).

## 4. *N*-fold rotationally symmetric ring of cores

In this section we develop expressions for the dispersion relation and the polarization state of the HBMs in a chiral 6-SRF and explore how they vary with twist rate. The normalized structural parameters used in the numerical simulations (see Fig. 3(a)) are the same as those given in section 3.2. In addition, FEM yields a normalized coupling constant (between two parallel untwisted waveguides) $\kappa/k = 0.00033$, which is used to derive the analytical results in the following sections. FEM also indicates that the coupling rate can, to a very good approximation, be considered independent of the polarization state for our parameters.

### 4.1 Dispersion relation of HBMs in Frenet-Serret frame

To treat the case of $N$ coupled birefringent cores arranged regularly spaced in a ring around the fiber axis (see Fig. 2(b)), we start with the expression in Eq. (10) and assume nearest-neighbor coupling between cores. Bloch's theorem requires that the field amplitudes and polarization states in neighboring cores be identical, when evaluated within the $(\sigma_1, \sigma_2)$ coordinate system, except for the Bloch phase advance, which progresses along curves of constant $\sigma_1$, i.e., in the $\sigma_2$ direction (Fig. 1).

At a given value of $\sigma_1$, we can without loss of generality specify the positions of three neighboring cores as $\sigma_2 = n\Lambda_B$, where $n = (-1, 0, +1)$ and $\Lambda_B = 2\pi\rho/(N\sqrt{1+\alpha^2\rho^2}) = \delta\phi\rho/\sqrt{1+\alpha^2\rho^2}$ is the intercore arc-length along $\sigma_1$ = constant curves and $\delta\phi$ is the azimuthal angle between adjacent cores in the Cartesian $z$ = constant plane. Using Eqs. (2) and (3), the complex unit vector in the $n$-th core can then be written to a very good approximation in the form:

$$\hat{u}_n(\sigma_1) = -\frac{(\hat{x}+i\hat{y})}{\sqrt{2}} e^{-i(\phi_T + n\delta\phi)} = \hat{u}_0(\sigma_1) e^{-in\delta\phi} \qquad (23)$$

where the small $z$-component has been neglected. The field Ansatz in the $n$-th core can then be written:

$$\mathbf{E}_n(\sigma_1) = \left(e_{Ln}(\sigma_1)\hat{u}_0(\sigma_1)e^{-in2\pi/N} + e_{Rn}(\sigma_1)\hat{u}_0^*(\sigma_1)e^{in2\pi/N}\right)e^{i\beta_0\sigma_1} \qquad (24)$$

where $\hat{u}_0(\sigma_1)$ is taken to be evaluated at $\sigma_2 = 0$. Bloch's theorem allows us to write:

$$e_{Ln}(\sigma_1) = e_{L0}(\sigma_1)e^{in\phi_B}, \quad e_{Rn}(\sigma_1) = e_{R0}(\sigma_1)e^{in\phi_B} \qquad (25)$$

where $\phi_B$ is the Bloch phase advance between adjacent cores, leading to:

$$\mathbf{E}_n(\sigma_1) = \left(e_{L0}(\sigma_1)\hat{u}_0(\sigma_1)e^{in(\phi_B - 2\pi/N)} + e_{R0}(\sigma_1)\hat{u}_0^*(\sigma_1)e^{in(\phi_B + 2\pi/N)}\right)e^{i\beta_0\sigma_1}. \qquad (26)$$

We are now in a position to set up the coupled mode equations. Substituting Eqs. (26) into Eq. (14), allowing the cores to couple, neglecting second order derivatives of the field amplitudes and taking the dot product with $\hat{u}$ and $\hat{u}^*$ yields:

$$\begin{pmatrix} \gamma - \tau_1 - 2\kappa\cos\left(\phi_B - \frac{2\pi}{N}\right) & -\frac{\vartheta}{2}e^{i2\phi_T} \\ -\frac{\vartheta}{2}e^{-i2\phi_T} & \gamma + \tau_1 - 2\kappa\cos\left(\phi_B + \frac{2\pi}{N}\right) \end{pmatrix} \begin{pmatrix} e_{L0} \\ e_{R0} \end{pmatrix} = 0 \qquad (27)$$

after making the same approximations as used in deriving Eq. (16). Eigenvalue analysis of this system yields for the propagation constant $\beta_T$ in the tangential direction:

$$\beta_T = \beta_0 + \gamma = \beta_0 + 2\kappa\cos\phi_B\cos\frac{2\pi}{N} \pm \sqrt{\frac{\vartheta^2}{4} + \left(2\kappa\sin\phi_B\sin\frac{2\pi}{N} + \tau_1\right)^2} \qquad (28)$$

which reduces for zero linear birefringence to:

$$\beta_T = \beta_0 \pm \tau_1 + 2\kappa\cos(\phi_B \mp 2\pi/N) \qquad (29)$$

for LCP and RCP modes, as follows from Eq. (27) by simple inspection.

*4.2 Group velocity of HBMs in Frenet-Serret frame*

Although as we shall discuss later only discrete values of $\phi_B$ are allowed (dictated by the need for an azimuthal resonance), the shape of the full dispersion surface (upper plot in Fig. 4) reveals the direction of the group velocity of the HBMs, which points normal to the dispersion surfaces in the direction of increasing optical frequency and for a given HBM is identical for every harmonic [31]. The relative magnitudes of the group velocity components in the $\hat{N}_1$ and $\hat{B}_1$ directions can be found by taking the gradient of the dispersion relation, expressed as a Hamiltonian [31]:

$$H = \beta_T - \left(\beta_0 + 2\kappa\cos(k_B\Lambda_B)\cos\frac{2\pi}{N} \pm \sqrt{\frac{\vartheta^2}{4} + \left(\tau_1 + 2\kappa\sin(k_B\Lambda_B)\sin\frac{2\pi}{N}\right)^2}\right) = 0, \qquad (30)$$

with respect to $\beta_T$ and $k_B$:

$$\mathbf{v}_G = \frac{c}{n_0}\left(\hat{T}_1 + \hat{B}_1 \partial H/\partial k_B\right)/\sqrt{1 + (\partial H/\partial k_B)^2}. \qquad (31)$$

The normalized group velocity component in the $\hat{B}_1$ direction is plotted in Fig. 4 (lower), showing that the fraction of power flowing in the azimuthal direction does not depend in a simple way on the azimuthal order, but rather is proportional to the weighted average of the power carried by the HBM harmonics, just as is the case for normal photonic Bloch waves [32].

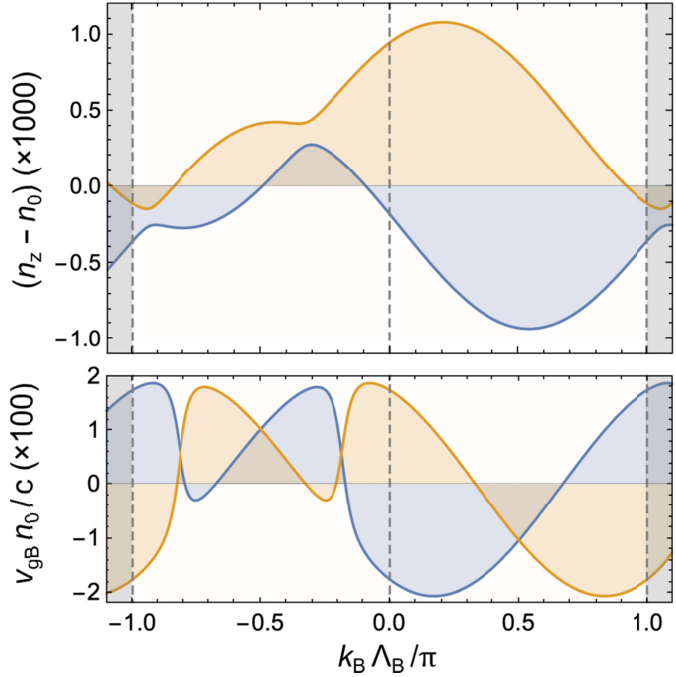

Fig. 4: Upper: Dispersion surfaces of the two HBMs in the Frenet-Serret frame, plotted against the Bloch phase for $N = 6$, $\vartheta/k = 0.00014$, $\alpha/k = 0.00033$, $k\rho = 9\pi$, $\kappa/k = 0.00057$ and $n_0 = 1.441$, where $k$ is the vacuum wavevector. Note that discretization of the modes, caused by the closed azimuthal path, is not included (see Fig. 5). Lower: Normalized group velocities in the $\hat{B}_1$ direction, $v_{gB}$, expressed as a fraction of the phase velocity $c/n_0$.

### 4.3 Amplitudes of HBM harmonics

As the HBMs propagate in the twisted fiber, their field pattern (formed by the interference of harmonics with different propagation constants) rotates in synchronism with the cores. The amplitudes $V_m$ of individual harmonics can be estimated by assuming that the azimuthal field profile in each core is relatively unperturbed by the inter-core coupling and the twist and so may be written $e_c(\phi) = e_c(\phi + 2n\pi/N)$ where $1 \leq n \leq N$ is the unit cell number [33]. Expanding $e_c(\phi)$ as a Fourier series with period $2\pi/N$, we obtain:

$$V_m = \frac{N}{2\pi} \int_{-\pi/N}^{\pi/N} e_c(\phi) \exp(-i\ell_A^{(m)} \phi) d\phi \tag{32}$$

which reduces to:

$$V_m = \frac{N \delta\phi_c}{2\pi} e_{c0} \operatorname{sinc} \frac{\ell_A^{(m)} \delta\phi_c}{2} \tag{33}$$

if we approximate the fields to constant (equal to $e_{co}$) within and zero outside the cores, which subtend an angle $\delta\phi_c$. If on the other hand we assume Gaussian field profiles with 1/e full width $\delta\phi_g$, we obtain approximately:

$$V_m = \frac{N \delta\phi_g}{2\sqrt{\pi}} e_{c0} \exp\left(-(\ell_A^{(m)} \delta\phi_g/2)^2\right). \tag{34}$$

Although these expressions neglect the radial dependence of the fields, they usefully highlight the azimuthal structure of the HBMs. In practice the amplitudes of the harmonics can be very significant, even out to quite high orders, as reported in recent experiments [28].

### 4.4 Transforming to the laboratory frame

The total wavevector, which in the Frenet-Serret frame is

$$\mathbf{k} = (\beta_0 + \gamma(k_B \Lambda_B))\hat{T}_1 + k_B \hat{B}_1, \quad k_B = \phi_B / \Lambda_B, \tag{35}$$

can be referred to the Cartesian $(x,y,z)$ frame by replacing $\sigma_1$ and $\sigma_2$ using Eq. (2), leading to:

$$\begin{aligned}
\beta_z &= (\beta_0 + \gamma)\hat{T}_1 \cdot \hat{z} + k_B \hat{B}_1 \cdot \hat{z} = (\beta_0 + \gamma - k_B \alpha \rho)/\sqrt{1 + \alpha^2 \rho^2} \\
\mathbf{k}_{az} &= (\beta_0 + \gamma)\hat{T}_1 + k_B \hat{B}_1 - \beta_z \hat{z} = \frac{k_B + \alpha\rho(\beta_0 + \gamma)}{\sqrt{1 + \alpha^2 \rho^2}}(-\sin\phi, \cos\phi, 0) \\
|\mathbf{k}_{az}| &= \left|\frac{k_B + \alpha\rho(\beta_0 + \gamma)}{\sqrt{1 + \alpha^2 \rho^2}}\right| = \left|\frac{\phi_B N}{2\pi\rho} + \frac{\alpha\rho(\beta_0 + \gamma)}{\sqrt{1 + \alpha^2 \rho^2}}\right|.
\end{aligned} \tag{36}$$

Applying the azimuthal resonance condition $k_{az}\rho = \ell_A$ and rearranging, we obtain for the $m$-th Bloch harmonic:

$$\phi_B = \frac{2\pi}{N}\left(\ell_A + mN - \alpha'\rho^2(\beta_0 + \gamma)\right) \simeq \frac{2\pi}{N}\left(\ell_A + mN - \alpha'\rho^2\beta_0\right) \tag{37}$$

and

$$\beta_z = (\beta_0 + \gamma)\sqrt{1+\alpha^2\rho^2} - \alpha(\ell_A + mN), \qquad (38)$$

which predicts that the dispersion surfaces tilt at an angle $\sin^{-1}(\alpha\rho/\sqrt{1+\alpha^2\rho^2})$ to the fiber axis. This is because they are locked to the $(\sigma_1, \sigma_2)$ coordinate system defined by the normal and binormal unit vectors.

Note that these expressions, rigorously derived, differ from previously reported heuristic results which did not take account of the torsion of the unit vectors as the mode propagates along the fiber [27]. Although this has negligible consequences for circularly polarized HBMs (the previous analyses did not take account of the small residual circular birefringence in Eq. (19)), it becomes important when the HBMs are elliptically or linearly polarized, as happens at the anti-crossing points where the RCP and LCP dispersion surfaces intersect.

## 5. Results

We now present the results of a series of calculations of the modal indices, polarization states, loss and field profiles of the HBMs at different twist rates, comparing with FEM results at each stage. We use the structural parameters listed in section 3.2 for a 6-SRF, with intercore coupling constant $\kappa/k = 0.00033$. Figure 5(a) shows a sketch of the fiber structure, consisting of six tilted elliptical cores in a ring.

### 5.1 Modal indices and polarization states of the HBMs

The modal index $n_z = \beta_z / k$ in the laboratory frame (Eq. (38)) is plotted versus azimuthal order in Fig. 5(b) for $N = 6$ and two values of twist rate. Note the anti-crossings, caused by linear birefringence in the cores, which couples LCP and RCP light. For $\alpha\lambda/2\pi = 0.00033$ they appear in the vicinity of $\ell_A = -2$ and 0. When the twist rate is increased to $\alpha\lambda/2\pi = 0.00057$ these two anti-crossings coincide at $\ell_A \sim -1$, cancelling each other out. Note that the refractive indices of the HBMs are unaffected by the tilt angle in both the analytical theory and the FEM.

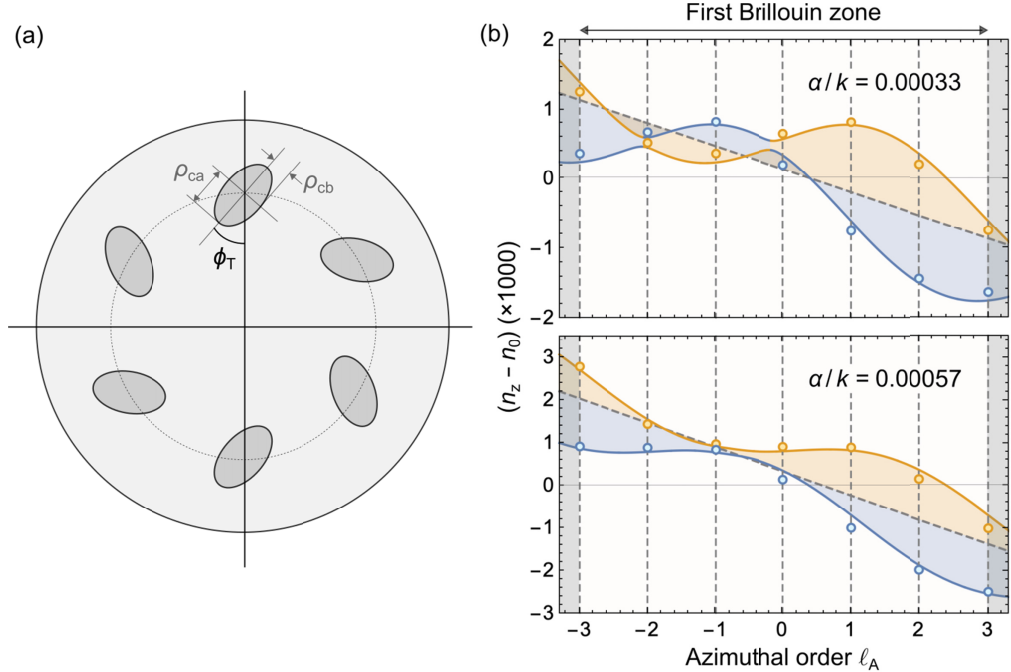

Fig. 5: (a) Sketch of a 6-SRF with core tilt angle $\phi_T$. (b) Modal indices within the first Brillouin zone in the laboratory frame, plotted against azimuthal order for twist rates $\alpha/k = 0.00033$ (upper) and $0.00057$ (lower). The other parameters are $k\rho_{ca} = 6.5\pi$, $k\rho_{cb} = 3.2\pi$, $k\rho = 9\pi$, $n_0 = 1.441$, $\vartheta/k = 0.00014$ and $\kappa/k = 0.00033$, where $k$ is the vacuum wavevector. The solid lines are the dispersion surfaces calculated using the analytical theory and the colored circles mark the indices of the HBMs, calculated by FEM. The colors mark modes with the same polarization state. Note the two anti-crossings that coincidentally appear in the vicinity of $\ell_A = -2$ and $0$ for $\alpha/k = 0.00033$. For $\alpha/k = 0.00057$ these anti-crossings coincide at $\ell_A \sim -1$, cancelling each other out.

The Stokes parameters of the HBMs in Fig. 5(b) are plotted versus azimuthal order in Fig. 6. Figures 6(a)-(c) explore the dependence on the tilt angle for $\alpha\lambda/2\pi = 0.00033$, showing a strong dependence. The analytical theory agrees very well with FEM in cases when the polarization state is close to circular, deviating only in the vicinity of the anti-crossings. One of these coincidentally occurs at $\ell_A = -2$, at which point the polarization state becomes perfectly linear with $S_1 = +1$ (Fig. 6(a)), in contrast to the $\ell_A = +2$ HBM, which is close to perfectly circularly polarized. For $\alpha\lambda/2\pi = 0.00057$ (Fig. 6(d)) the $\ell_A = -1$ HBM is close to linearly polarized, while the $\ell_A = +1$ HBM is circularly polarized. Considering that the fundamental modes of a circularly symmetric core have parameters $(\ell_A, s) = \pm(1, -1)$ (where $s = +1$ for LCP and $-1$ for RCP), this opens up interesting possibilities for novel kinds of polarization optics. A detailed comparison of the results of the analytical theory with FEM is available in section 6.2.

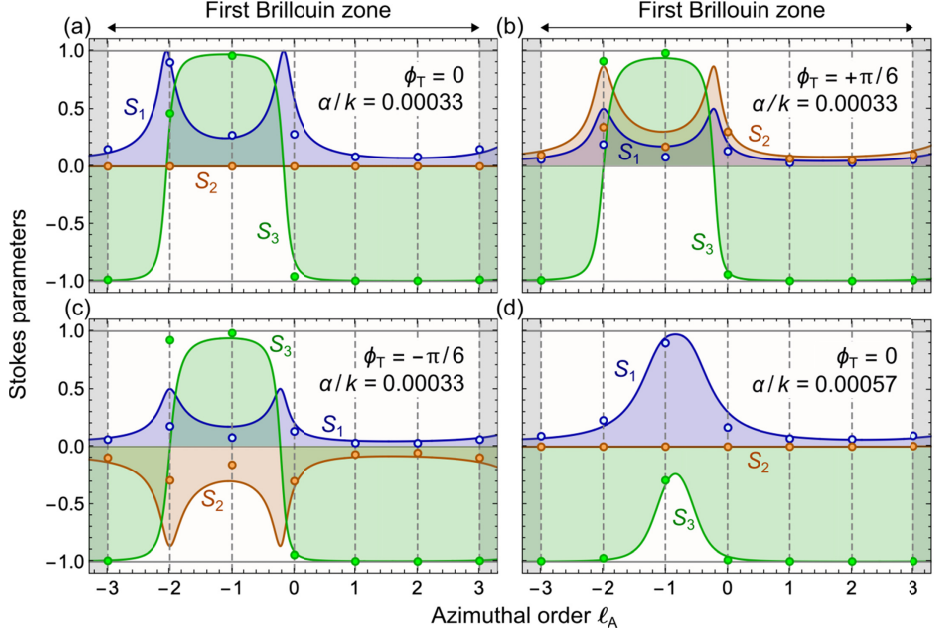

Fig. 6: Stokes parameters $S_m$ in cylindrical coordinates plotted versus azimuthal order $\ell_A$ for the lower dispersion surface in Fig. 5(b). $S_3 = \pm 1$ corresponds to circular polarization. For the other dispersion surface the Stokes parameters have identical magnitudes but opposite signs. The encircled data-points are calculated by FEM, and show quite good agreement with the analytical theory (full curves). (a-c) Plots for twist rate $\alpha/k = 0.00033$ at three different tilt angles $\phi_T$. (d) Plot for $\alpha/k = 0.00057$ and $\phi_T = 0$. In this case $S_3$ deviates strongly from $-1$ close to the double anti-crossing (see the lower plot in Fig. 5(b)).

*5.1 Twist rate dependence*

The dependence of the HBM refractive index on twist rate is of key interest in many applications. The indices in the laboratory frame of the $\ell_A = 0, +1, +2,$ and $+3$ HBMs in the first Brillouin zone are plotted in Fig. 7(a) and (b) as a function of normalized twist rate $\alpha/k$. As the twist rate increases, the dispersion surface tilts more and more, higher values of $\ell_A$ experiencing the strongest twist-dependence. In some cases the index first falls and then rises as the twist rate increases. This is caused by the sinusoidal shape of the dispersion surfaces. The Stokes parameters of the HBMs in the local Frenet-Serret frame are plotted in Fig. 7(c) and (d). The plots for negative values of $\ell_A$ may be obtained by reflecting the diagrams in Fig. 7 about the vertical axis.

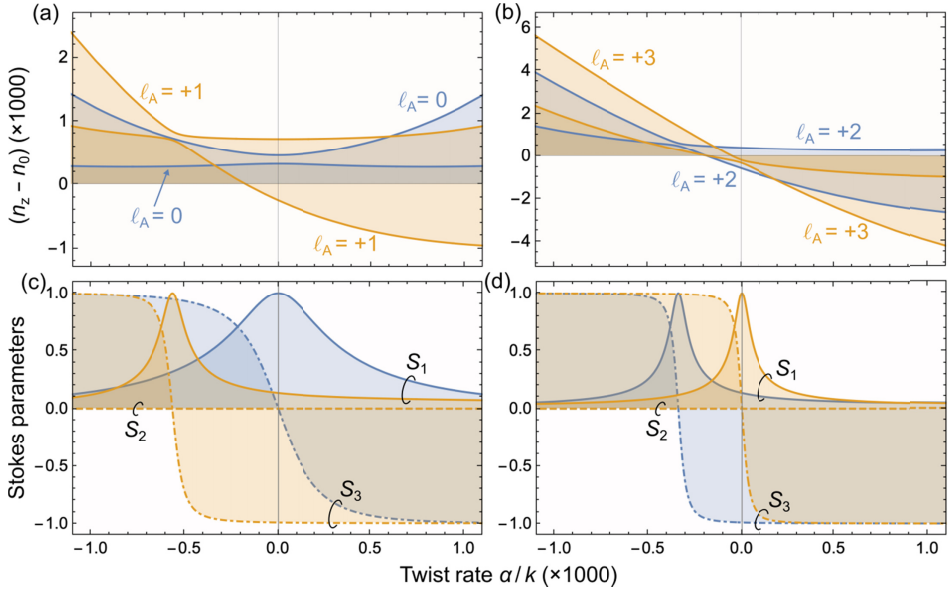

Fig. 7: HBM parameters with increasing twist rate $\alpha$ for $N = 6$ and the same parameters as in Fig. 5. (a-b) Refractive index $n$ of the HBMs in the first Brillouin zone in the laboratory frame. For negative values of $\ell_A$ the curves are reflected about the $\alpha\lambda/2\pi = 0$ axis. (c-d) Stokes parameters $S_1$, $S_2$ and $S_3$ for the lower curves in (a) and (b), calculated in the local Frenet-Serret frame. For the upper curves the Stokes parameters have the same magnitude but the opposite sign.

*5.2 Azimuthal order, spin and orbital angular momentum*

The angular momentum order $\ell_{AM}$ of a circularly polarized Laguerre-Gaussian beam is an integer, being linked to the azimuthal order $\ell_A$ (in scalar analyses often called the orbital angular momentum order) and spin by $\ell_{AM} = \ell_A - s$. In more complex situations, for example in a $C_N$ fiber structure, this is only true under special circumstances, since the polarization state of the light is not always circular. The azimuthal order, on the other hand, is always a good quantum number no matter how complex the structure, i.e., it is always an integer [27]. This means in turn that the field amplitudes of any HBM, expressed in the local Frenet-Serret frame, repeat perfectly $N$ times around the azimuth, after the Bloch phase factor is removed.

For zero linear birefringence in the cores ($\vartheta = 0$), the HBMs are circularly polarized and spin can be incorporated directly into the dispersion relation via spin-orbit coupling. Under these circumstances Eq. (38) yields, after some manipulation:

$$\frac{\beta_z - \beta_0}{\sqrt{1+\alpha^2\rho^2}} = s(\tau_1 - \alpha') - \alpha'\ell_{AM}^{(m)} + \kappa\cos\left(2\pi(\ell_{AM}^{(m)} + \alpha'\rho^2\beta_0)/N\right) \qquad (39)$$

where $\ell_{AM}^{(m)} = \ell_A^{(m)} - s$. Since the first term on the right is very small for our parameters (see discussion around Eq. (19)), Eq. (39) shows that the dispersion surfaces depend, to a very good approximation, solely on the angular momentum order, collapsing on to one curve. When the cores are linearly birefringent, however, this collapse is not possible because clean spin-orbit coupling does not in general exist, especially close to the anti-crossings where the linear birefringence couples LCP and RCP fields. Note, however, that when the field is close to perfectly circularly polarized (e.g., for $\ell_A = 0, +1$ and $+2$ in Fig. 6), Eq. (39) is approximately valid. This is illustrated in Fig. 8 for the case of perfect circular polarization states.

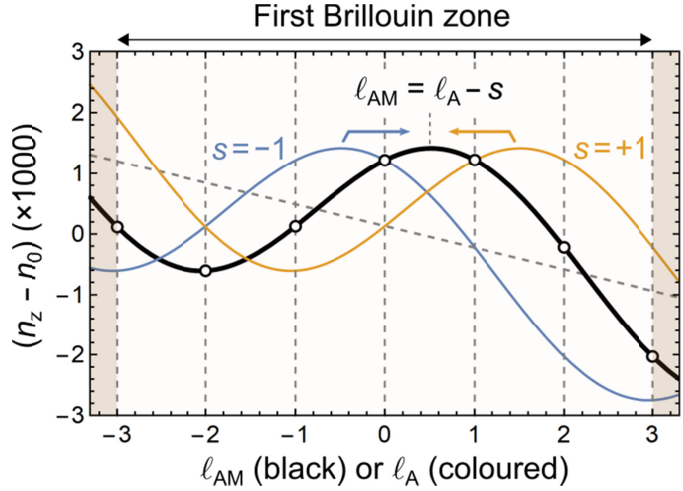

Fig. 8: Dispersion surfaces for circularly polarized HBMs in the laboratory frame in the absence of linear core birefringence ($\vartheta = 0$). The thin colored curves are plotted against azimuthal order and the thick black curve against the total AM order $\ell_{AM} = \ell_A - s$. The absence of an anti-crossing between the azimuthal curves means that the two HBMs are uncoupled, so that the azimuthal order can be converted to AM order $\ell_{AM}$ simply by subtracting the spin $s$, resulting in two almost identical curves for RCP and LCP modes (the residual circular birefringence, originating from $\tau_1 - \alpha' \neq 0$, is much less than the thickness of the black curve).

*5.3 Optomechanical torque*

It is interesting at this point to ask what optomechanical torque will act on the twisted fiber when a HBM is excited. This is related to the mean azimuthal momentum of the HBM, which can be expressed as a weighted sum over all the harmonics:

$$\langle \ell_A \rangle = \frac{\sum_m |V_m|^2 \ell_A^{(m)}}{\sum_m |V_m|^2} \qquad (40)$$

and is closely related to the group velocity component in the azimuthal direction. The spin angular momentum of the light can in principle be added to this summation, but note that the light is not in general perfectly circularly polarized. The mechanical torque resulting solely from Eq. (40) is then given by $\langle \ell_A \rangle P/\nu$, where $P$ and $\nu$ are the optical power and frequency.

## 6. Comparison of analytical results with FEM

In this section we compare the modal parameters calculated analytically with those obtained by FEM, and show that the disparities are caused mainly by twist-dependent distortions in the modal field distributions. FEM also allows estimation of the leakage loss caused by the twist.

### *6.1 Modal field profiles*

In Fig. 9 the field intensity profiles of the HBMs, calculated by FEM for the same parameters as in Fig. 5, are plotted for $\alpha/k = 0.002$, $0.004$, and $0.005$. We have already seen in section 3.2 that the field profiles in a single-core fiber become distorted as the twist rate increases, first moving outwards within the elliptical core (a centrifugal effect caused by the curvature), becoming smaller and more circular, and then above a certain critical twist rate leaking strongly into the cladding. These effects will also occur for HBMs, with the additional consequence of a reduction in coupling between the cores as the twist rate increases.

For the two lower twist rates the profiles are almost identical for HBMs with the same total angular momentum order $\ell_{AM}$ but opposite spin. As $|\ell_{AM}|$ increases, however, the mean radial position of the mode moves outwards, as one might extent from centrifugal considerations (Fig. 9), and the leakage loss rises. At $\alpha/k = 0.005$ modes with the same total angular momentum but opposite spin have slightly different profiles, those with $S_3 = +1$ moving to a slightly higher mean radius. The magnitude of $S_3$ is within 15% of unity in all the cases shown, since the anti-crossings occur only at lower values of twist (see Fig. 7).

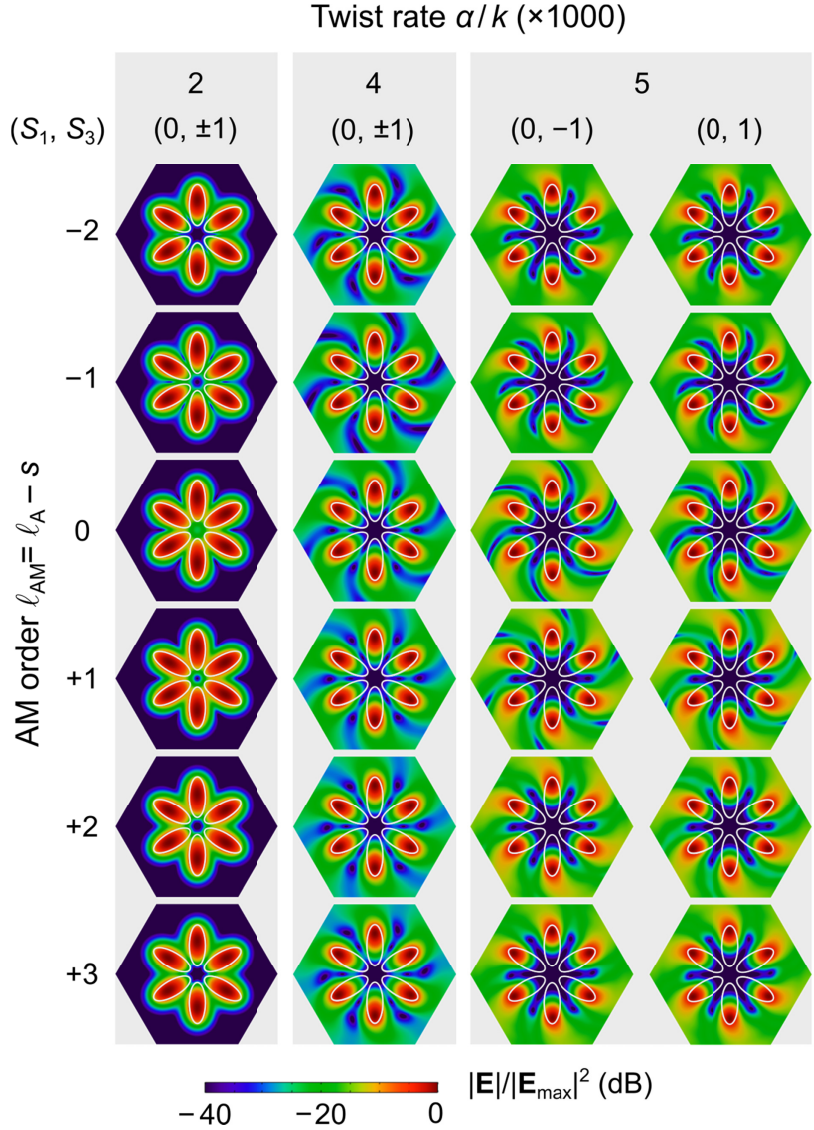

Fig. 9: Electric field intensity distributions of the HBMs in a 6-SRF fiber for three different twist rates $\alpha/k = 0.002, 0.004$ and $0.005$. At the two lower twist rates the field distributions are almost identical for HBMs with the same total AM order $\ell_{AM} = \ell_A - s$ but orthogonal polarization states, so only one distribution is shown. At the highest twist rate some small differences appear between HBMs with the same OAM order but opposite polarization states, so both are shown. The polarization states are very close to perfectly circular at these relatively high twist rates (anti-crossings occur only at lower values of twist rate, as seen in Fig. 6). Note that the values given for the Stokes parameters (at the top of each column) are approximate, with errors below ~15% in all cases.

## 6.2 Modal indices, Stokes parameters and loss

In Fig. 10 the differences between the analytical and FEM values of the HBM refractive indices and Stokes parameters are plotted versus twist rate for the same parameters as in Fig. 5. It is remarkable how the deviation of $S_3$ from $\pm 1$ (green curves) is very small in all cases except in the vicinity of the anti-crossings. The largest errors occur for the $S_1$ and $S_2$ parameters, mostly remaining below ±1% over the twist range −2 < α/k < +2 but reaching ±6% in a few cases, in the vicinity of anti-crossings.

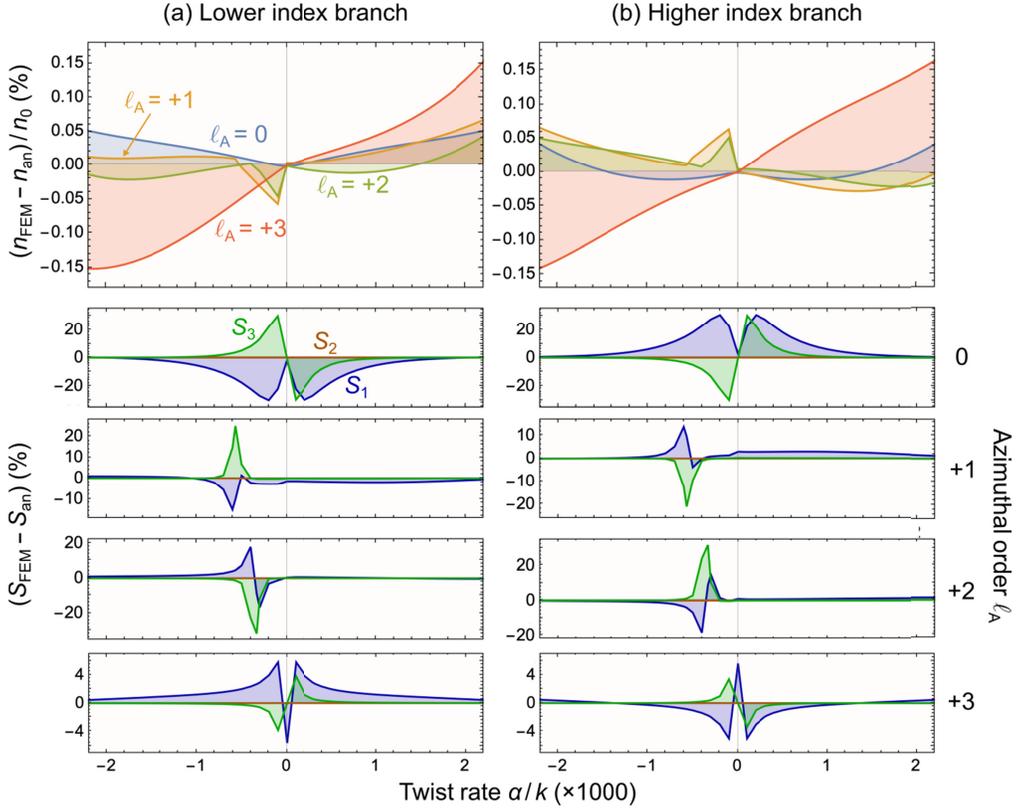

Fig. 10: Differences between modal values calculated analytically (subscript "an") and those obtained by FEM. The index difference is expressed as a percentage of $n_0$, and the Stokes parameter errors as percentages of unity.

## 7. Conclusions

Fully vectorial coupled-mode equations based in the local Frenet-Serret frame, presented here for the first time, provide a convenient system for analyzing the properties of helical Bloch modes, which appear in *N*-fold rotationally symmetric rings of coupled cores. Cores that are birefringent and tilted can be treated. Agreement between the analytical solutions and FEM is very good at low twist rates, when the modal field profiles remain relatively undistorted. At higher twist rates, centrifugal effects push the HBMs out radially and alter the birefringence and the inter-core coupling rate, reducing the agreement between analysis and FEM. This disagreement could be corrected by using FEM to find the correct values for the effective radial position, birefringence and coupling rate at each twist rate and entering these into the analytical expressions. At a fixed twist rate, the FEM mode profiles are observed move slightly outwards as the total angular momentum increases, once again as a result of centrifugal effects.

    The analysis can be used to treat more complicated $C_N$ structures of coupled cores by first finding the HBMs for cores positioned equi-radially and then permitting them to couple. Note, however, that coupling between HBMs at different values of radius is only possible if the azimuthal orders are identical and the polarization states non-orthogonal. One important

example, explored in many papers, concerns coupling between a central on-axis core and a surrounding chiral microstructure [18] [24] [34]. In the special case of an N-SRF with a central (single-mode and birefringence-free) core, coupling is only possible between core and ring if $\ell_A = \pm 1$ and the polarization states are at least partially non-orthogonal [27]. This is because the central core can only support modes with total angular momentum order $\ell_{AM} = 0$, i.e., azimuthal order and spin $(\ell_A, s) = \pm(1, -1)$. The same general considerations will apply if the central core is multimode.

In conclusion, vector coupled-mode theory based in the local Frenet-Serret frame provides an elegant and accurate means of analyzing and understanding the properties of helical Bloch modes in twisted N-fold rotationally symmetric single-ring fibers. We believe it will provide a firm basis for future studies of more complex multi-core structures, both linear and nonlinear.

**Disclosures.** The authors declare no conflicts of interest.